\documentclass{Interspeech}
\usepackage{pifont,amsmath,amssymb,caption,subcaption}
\newcommand{\xmark}{\ding{55}}

\interspeechcameraready

\title{Score-Based Training for Energy-Based TTS Models}

\author[affiliation={1}]{Wanli}{Sun}
\author[affiliation={1}]{Anton}{Ragni}

\affiliation{School of Computer Science}{University of Sheffield}{Sheffield, UK}

\email{wsun20@sheffield.ac.uk, a.ragni@sheffield.ac.uk}

\usepackage{comment}
\usepackage{pgfplots}
\usepgfplotslibrary{groupplots}
\pgfplotsset{compat=1.18}

\begin{document}

\maketitle

\begin{abstract}
    
    Noise contrastive estimation (NCE) is a popular method for training energy-based models (EBM) with intractable normalisation terms. The key idea of NCE is to learn by comparing unnormalised log-likelihoods of the reference and noisy samples, thus avoiding explicitly computing normalisation terms. However, NCE critically relies on the quality of noisy samples. Recently, sliced score matching (SSM) has been popularised by closely related diffusion models (DM). Unlike NCE, SSM learns a gradient of log-likelihood, or \textit{score}, by learning distribution of its projections on randomly chosen directions. However, both NCE and SSM disregard the form of log-likelihood function, which is problematic given that EBMs and DMs make use of first-order optimisation during inference. This paper proposes a new criterion that learns scores more suitable for first-order schemes. Experiments contrasts these approaches for training EBMs.

\end{abstract}

\section{Introduction}
Recently, energy-based models\,(EBMs)~\cite{lecun2006tutorial, wenliang2019learning} have drawn some attention in the text-to-speech\,(TTS) area~\cite{sun2024energy}. These probabilistic models define the log-likelihood of speech given text as the difference between negative (unnormalised) energy function and the logarithm of the corresponding normalisation term.
Due to the difficulty of computing normalisation terms, noise contrastive estimation\,(NCE) has been previously utilised to train EBMs in applications such as speech recognition~\cite{li2021residual}, machine translation~\cite{bhattacharyya2020energy}, text generation~\cite{bakhtin2021residual}, and TTS \cite{sun2024energy}. The NCE compares unnormalised log-likelihoods of `positive' (reference) examples to unnormalised log-likelihoods of `negative' (noisy) samples to decide the direction for updating model parameters. Unlike language modelling, where NCE appears to work well with a wide choice of negative samples~\cite{chen2015recurrent}, TTS appears to be much more sensitive to the quality of negative samples and necessiates the development of elaborate sampling schemes~\cite{sun2024energy}. 

Several alternative approaches have been proposed in the literature: score matching\,(SM)~\cite{hyvarinen2005estimation}, denoising score matching\,(DSM)~\cite{vincent2011connection}, and sliced score matching\,(SSM)~\cite{song2020sliced}. All these alternatives attempt to predict the gradient of log-likelihood with respect to data (speech in TTS), or {\em score}. 
However, the use of SM would require computing the trace of the Hessian, which is expensive for automatic differentiation packages~\cite{song2021train}. 
The introduction of noise to data in DSM training leads to inconsistency between inference and training~\cite{song2021train}. 
Instead of matching scores, SSM matches their projections on randomly chosen directions, thus avoiding much expensive computation. SSM has been used with EBMs in many domains, e.g., imitation learning~\cite{kim2021imitation} and out-of-distribution detection~\cite{elflein2021out,liu2022effortless}, but has not yet been investigated in TTS. 

Score-based training of EBMs and currently popular diffusion style models~\cite{song2019generative, liu2022flow, kim2020glow} is typically followed by a variant of Markov Chain Monte-Carlo\,(MCMC) for performing inference. As a type of first-order optimisation, MCMC-style approaches are sensitive to the nature of objective function they optimise. For example, a highly non-convex shape may lead to issues in converging to an acceptable solution within acceptable time. Despite this, score-based approaches appear to focus on learning gradients as accurately as possible, regardless of the shape of the underlying objective. This paper introduces a new score-based learning objective, called {\em delta loss}, that penalises those scores that do not lie on linear paths between current speech estimates and reference speech samples. As such, delta loss is believed to be more appropriate for score-based models that utilise MCMC-style first-order optimisation approaches for inference. This paper also shows that although differently motivated the new loss function is closely linked with training objectives of currently popular flow matching approaches~\cite{liu2022flow}.

This paper makes the following key contributions:
\begin{enumerate}
\item first application of SSM to training EBMs in TTS;
\item a new score-based loss function for EBMs;
\end{enumerate}
and the rest of it is organised as follows. Section~\ref{sec:ebm} discusses EBMs and related work, including training methods and connections between EBMs and diffusion models. Section~\ref{sec:score} describes existing score-based training approached and introduces delta loss as a possible alternative. Experimental results and discussion are presented in Section~\ref{sec:exp}. Conclusions drawn from this work are presented in Section~\ref{sec:end}.

\section{Energy-Based Models}
\label{sec:ebm}
Given a sequence of text tokens $\bm x$, an energy-based model (EBM) of spectral feature sequences $\bm Y$ could be defined as \cite{sun2024energy}
\begin{equation}
p_{\bm \theta}(\bm{Y} | \bm{x})=\frac{1}{Z_{\bm \theta}(\bm {x})} \exp (-E_{\bm \theta}(\bm{x}, \bm{Y})),
\label{EBM}
\end{equation}
where $E_{\bm \theta}(\bm x, \bm Y)$ is an energy function with parameters $\bm \theta$ that connects text and speech such that smaller values correspond to better matches between text and speech and {\em vice versa}, and $Z_{\bm \theta}(\bm x)$ denotes the normalising term, which is intractable to calculate. Although intractable, normalising terms cancel out in log-likelihood ratios, which is the fact exploited in Section~\ref{ssec:nce}, or disappear in the following gradients of log-likelihood
\begin{equation}
\begin{split}
    \nabla_{\bm{Y}} \log p_{\bm{\theta}}(\bm Y |\bm{x}) = -\nabla_{\bm{Y}} E_{\bm{\theta}}(\bm{x}, \bm Y) -\underbrace{\nabla_{\bm{Y}} \log Z_{\bm{\theta}}(\bm x)}_{=0}, 
\end{split}
\label{energyWithoutZ}
\end{equation}
which is the fact exploited in Section~\ref{sec:score} and by gradient-based inference schemes. In its simplest form, inference with EBMs could be performed using the following version of Langevin Markov Chain Monte-Carlo\,(MCMC)~\cite{sun2024energy}
\begin{equation}
  {\bm Y}^{(N+1)} \leftarrow {\bm Y}^{(N)} + \rho \nabla_{\bm Y} \log p_{\bm \theta}(\bm{Y}^{(N)}|\bm{x}),
\label{langevin_mcmc}
\end{equation}
where 
$\rho$ is an learning rate. Typically, a large number of iterations is required to yield accurate speech samples ~\cite{sun2024energy}, which is consistent with other related models~\cite{song2019generative, song2020score, ho2020denoising}. 

\subsection{Noise contrastive estimation}
\label{ssec:nce}
Because of the intractable normalising term $\bm Z_{\bm\theta}(\bm x)$, training approaches that do not involve normalising terms, such as noise contrastive estimation\,(NCE)~\cite{gutmann2010noise}, could be adopted. As its name suggests, learning in NCE is performed by contrasting positive samples ${\bm Y}^{+}$ with negative, noisy, samples ${\bm Y}^{-}$. Specifically, the loss function in NCE is given by
\begin{equation}
    \begin{split}
         \mathcal{L}_{\bm\theta}^{\tt nce}({\bm x},{\bm Y}^{+},{\bm Y}^{-}) = &-\log\left( \frac{1}{1+\exp \left(E_{\boldsymbol{\theta}}(\bm{x}, \bm{Y}^+)\right)}\right) \\
        &- \log \left(\frac{1}{1+\exp \left(-E_{\bm{\theta}}(\bm{x},  \bm{Y}^-)\right)}\right),
    \end{split}
    \label{nce loss function}
\end{equation}
where ${{\bm Y}^{+}}$ is a reference speech and ${{\bm Y}^{-}}$ is a noisy speech. According to equation~\eqref{nce loss function}, optimal energy functions assign high energies to noisy samples and low energies to positive samples. Although NCE does not prescribe which noisy samples to use, the previous work in TTS found it sensitive to the quality of noisy speech \cite{sun2024energy}. This appears to be different to language modelling where simple negative sampling approaches were found to lead to better results~\cite{chen2015recurrent}. Although high-quality noisy speech could be obtained by combining outputs from well-trained TTS models \cite{shen2018natural, li2019neural, ren2020fastspeech, mehta2024matcha} and data augmentation methods~\cite{sun2024energy}, such an approach would lead to an increased training cost.


\subsection{Scores}
\begin{figure}[t]
  \includegraphics[width=\linewidth]{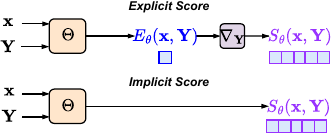}
    \caption{Two ways of computing \textit{scores} for EBMs: analytic (top) and predictive (bottom)}
    \label{fig:explicit&implicit}
\end{figure}
The gradient of log-likelihood with respect to speech $\nabla_{\bm Y} \log p_{\bm \theta}({\bm Y}|{\bm x})$ is known in the literature as a {\em score} ${\bm S}_{\bm\theta}({\bm x},{\bm Y})$. Such scores appear not only in EBMs but also in closely related diffusion models where the following inference rule may be used to denoise speech
\begin{equation}
  {\bm Y}^{(N+1)} \leftarrow \frac{1}{\sqrt{\alpha_N}} \{\bm Y^{(N)}+\beta_{N} {\bm S}_{\bm \theta}(\bm x, \bm Y^{(N)}, N)\} + \sigma_N \bm Z^{(N)},
  \label{denoising}
\end{equation}
where ${\bm S}_{\bm \theta}(\bm x, \bm Y^{(N)}, N)$ denotes a time (iteration) dependent score, $\bm Z \sim \mathcal{N}(\mathbf{0}, \mathbf{I})$, and $\alpha_N$, $\beta_N$, and $\sigma_N$ are hyperparameters. 
The close connection between EBMs in equation~\eqref{langevin_mcmc} and DMs becomes apparent by noticing that $\frac{1}{\sqrt{\alpha_N}}$ is a scale that combines iteration dependent weight decay parameters, $1-\frac{1}{\sqrt{\alpha_{N}}}$ with $1$ from the current speech estimate. Thus, setting $\alpha_N$ to $1$ and $\sigma_N$ to 0 (removes stochastic term) would lead to the update rule in equation~\eqref{langevin_mcmc}. 

In addition to scores not depending on iteration index in EBMs (${\bm S}_{\bm\theta}({\bm x},{\bm Y})$ vs ${\bm S}_{\bm\theta}({\bm x},{\bm Y},N)$), another interesting aspect of EBM scores is that they can either be modelled directly as in DMs or obtained by computing the gradient of log-likelihood. Figure~\ref{fig:explicit&implicit} illustrates these two different approaches. Although more expensive the latter approach ensures that scores are true gradients of some true log-likelihood function. 

\section{Score-based Training}
\label{sec:score}
This section will first discuss score matching approaches and then will introduce a new criterion, which is shown to be linked with currently popular flow matching~\cite{liu2022flow,lipman2022flow, heitz2023iterative}, that addresses some of score matching limitations. 

\subsection{Sliced score matching (SSM)}
Rather than using NCE it is also possible to optimise EBMs by matching their scores to those of the true data distribution. If the true data distribution, $p_{\tt data}({\bm x}, {\bm Y})$ was known then EBM training could be performed by minimising the distance between model score ${\bm S}_{\bm\theta}({\bm x},{\bm Y})$ and data score $\nabla_{\bm Y} \log p_{\tt data}({\bm x}, {\bm Y})$. As the latter is seldom known, an equivalent, practically feasible, formulation that does not depend on data scores has been found by~\cite{hyvarinen2009estimation}. It's unbiased estimator is given by
\begin{equation}
{\mathcal L}_{\bm\theta}^{\tt sm}({\bm x},{\bm Y}^{\tt +}) = \text{tr} \left( \nabla_{{\bm Y}^{+}} {\bm S}_{\bm\theta}({\bm x},{\bm Y}^{+}) \right) + \frac{1}{2} \Vert {\bm S}_{\bm\theta}({\bm x},{\bm Y}^{+})\Vert_{2}^{2},
\label{eq:sm}
\end{equation}
where the first term is the trace of the Hessian matrix, which is computationally expensive for high-dimensional data~\cite{song2020sliced} even with modern hardware~\cite{martens2012estimating}. Hence, direct score matching (SM) is only suitable for simple EBMs.

Fortunately, a computationally efficient variant of SM called sliced score matching (SSM) \cite{song2020sliced} has been proposed and could be adopted with complex EBMs. 
Given a random projection vector $\bm v$, the loss function of SSM is given by:
\begin{equation}
{\mathcal L}_{\bm\theta}^{\tt ssm}({\bm x}, {\bm Y}^{+}) = \bm{v}^{\top} \nabla_{\bm{Y}^{+}} \bm{S}_{\bm \theta}(\bm{x}, {\bm Y}^{+}) \bm{v}
+\frac{1}{2}\Vert\bm{S}_{\bm \theta}(\bm{x}, {\bm Y}^{+})\Vert_2^2.
\label{SSM loss}
\end{equation}
Computing the first term in equation~\eqref{SSM loss} requires only 2 gradient operations rather than $\dim({\bm Y}^{+})$ gradient operations required in equation~\eqref{eq:sm}. Although it is possible to use $K$ random projection rather than one, the single random projection version above is used in this work. Unlike NCE, which requires special care to crafting negative samples, SSM appears to be a straightforward, albeit an expensive, training approach. 

\subsection{Delta loss (Delta/$\Delta$)}
\begin{figure}[!ht]
\centering
\includegraphics[width=\linewidth]{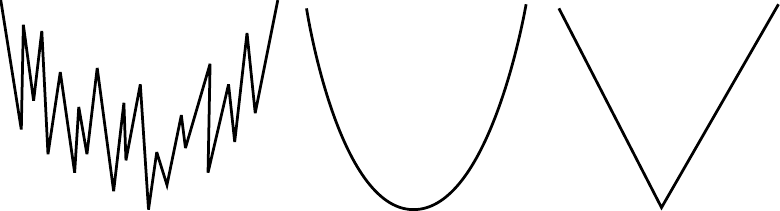}
\caption{Different score functions: from left to right, the hypotheses list from the worst to the best.}
\label{fig:losses shapes}
\end{figure}

Apart from requiring $1+K$ gradient computations, another drawback of SM-style approaches is the focus on matching scores without paying much attention to the underlying log-likelihood. Consider, for example, the log-likelihood on the left-hand side of Figure~\ref{fig:losses shapes}. 
Even if (S)SM above would manage to perfectly match data scores with model scores it is unlikely that first-order optimisation schemes such as equation~\eqref{denoising} would succeed in reaching its global minimum. Log-likelihoods such as those in the middle of Figure~\ref{fig:losses shapes} are friendlier for first-order schemes though might require careful tuning and large number of iterations. Finally, log-likelihoods on the right-hand side of Figure~\ref{fig:losses shapes} are optimal as they allow reaching global minima in one step provided correct learning rate has been selected. 
This suggests that pointing scores of noisy samples ${\bm Y}^{-}$ generated by an acoustic model in the direction of $\Delta({\bm Y}^{+},{\bm Y}^{-})={\bm Y}^{+}-{\bm Y}^{-}$ would yield model scores capable of reaching optima (${\bm Y}^{+}$) in one step of simple first-order inference scheme
\begin{equation}
{\bm Y}^{+} = {\bm Y}^{-} +  {\bm S}_{\bm\theta}(\bm{x}, \bm{Y}^{-})
\label{one step langevin_mcmc}
\end{equation}
using any noisy speech sample as initialisation. Such an objective function called {\em delta loss} in this paper is given by
\begin{equation}
\mathcal{L}_{\bm\theta}^{\Delta}({\bm x},{\bm Y}^{+},{\bm Y}^{-}) = \frac{1}{2}\Vert{\bm S}_{\bm \theta}(\bm x, {\bm Y}^{-}) - ({\bm Y}^{+} - {\bm Y}^{-})\Vert_2^2.
\label{delta loss function}
\end{equation}
Compared to both NCE and (S)SM, delta loss is a very efficient training approach.

\subsection{Connection to flow matching}
Interestingly, the delta loss in equation~\eqref{delta loss function} can be linked to the Flow Matching (FM) loss~\cite{liu2022flow,lipman2022flow, heitz2023iterative}
\begin{equation}
{\mathcal L}_{\bm\theta}^{\tt fm}({\bm x},{\bm Y}^{+},{\bm Y}^{0},t) = \frac{1}{2} \Vert {\bm V}({\bm x},\bm Y_{t}, t) - ({\bm Y}^{+} - {\bm Y}^{0})||^2,
\label{rectified flow loss} 
\end{equation}  
where $t \in[0,1]$, ${\bm Y}^0$ is a source or anchor sample possibly influenced by ${\bm x}$, $\bm Y_t=t \bm Y^{+}+(1-t) \bm Y^{0}$ and $\bm V(\bm x, \bm Y_t, t)$ is a function that transfers samples ${\bm Y}_{t}$ between source/anchor and data distributions. Despite their similarity, the FM loss is motivated by the slow inference of diffusion style approaches \cite{liu2022flow}, whereas delta loss is concerned with model scores being appropriate for first-order inference schemes. 

\section{Experiments}
\label{sec:exp}
\subsection{Experimental setup}
\subsubsection{Dataset}
The dataset utilized in this work is LJSpeech~\cite{ljspeech17}, which consists of 13,100 audio clips totalling approximately 24 hours of speech from a single female speaker reading passages from public domain texts. The dataset is randomly divided into training (10,000 clips), validation (1,800 clips), and test (1,300 clips) sets. A randomly chosen subset of 600 validation set clips was used in some objective evaluations to streamline the process. Another randomly chosen subset of 100 test set clips was used in subjective evaluations.

\subsubsection{Models}
The pretrained acoustic model generating hypotheses of spectral features is an open-source Tacotron 2\footnote{\scriptsize\url{https://github.com/NVIDIA/tacotron2}}. These spectral features are used as-is as an initialisation of inference for all EBMs in this paper. For training these hypotheses are either not used (SSM), or adopted as-is ($\Delta$) or additionally transformed to yield higher quality noisy samples (SSM). To ensure that references and hypotheses in training have the same duration (length), the pretrained acoustic model is forced to output spectral feature sequences with reference durations. No such constraint is enforced in inference. The non-score-based EBM trained with NCE is adopted from \cite{sun2024energy}. The score-based EBMs are based on the widely popular U-net architecture~\cite{ronneberger2015u}. The specific architecture is identical to Matcha-TTS~\cite{mehta2024matcha}\footnote{\scriptsize\url{https://github.com/shivammehta25/Matcha-TTS}} other than time/iteration was removed to yield time-independent scores. Each score-based EBM was trained for 50K steps using the Adam optimizer with a learning rate of $10^{-4}$ and a batch size of 10 on a single NVIDIA 3090 GPU. During inference, the score model is updating using a gradient descent algorithm with a learning rate of $3 \times 10^{-5}$. An open-source implementation of HiFi-GAN~\cite{kong2020hifi}\footnote{\scriptsize\url{https://github.com/jik876/hifi-gan}} is used as the vocoder to generate waveforms. 

\subsubsection{Evaluation}
\label{sec:evaluation}
The Mel cepstral distortion (MCD), log-scale F0 root mean square error (log f0), SpeechBERTScore~\cite{saeki2024speechbertscore}, and UTMOSv2~\cite{baba2024t05} are used as objective evaluation metrics. 
While the former two metrics are well known in the literature, the latter two, SpeechBERTScore and UTMOSv2, are model-based metrics that were previously found to be highly correlated with human perception. An implementation of these objective evaluation metrics is available in the {\footnotesize\tt DiscreteSpeechMetrics} toolkit\footnote{\scriptsize\url{https://github.com/Takaaki-Saeki/DiscreteSpeechMetrics}}.
Mean Opinion Score (MOS) is used as a subjective evaluation metric to assess the perceived naturalness of synthesized and reference speech. Five predictive models took part in listening tests (1xTacotron 2, 2xSSM, 2xDelta). Three listeners, who are all native English speakers, were asked to rate the perceived naturalness of each waveform on a scale from 1 (bad) to 5 (excellent) on the Amazon Mechanical Turk platform, with the average score representing the overall perceived naturalness.



\subsection{Objective evaluation}
Table~\ref{tab:loss} compares the performances of EBMs trained using SSM loss, delta loss or SSM and delta loss to the pretrained acoustic model, Tacotron 2. 
\begin{table}[htbp]
  \centering
  \begin{tabular}{c c|
  c c c c}
    \toprule
    \textbf{\scriptsize SSM} &
    \textbf{\scriptsize Delta} &
    \textbf{\scriptsize MCD $\downarrow$} & 
    \scriptsize $\log f{\mathrm{0}} \downarrow$ &
    \textbf{\scriptsize S-BERT $\uparrow$} &
    \textbf{\scriptsize{UTMOSv2 $\uparrow$} } \\
    \midrule
      \checkmark & \xmark & $\bm{5.312}$& $\bm{0.296}$& $\bm{0.905}$& 3.710\\
      \xmark & \checkmark & 5.399& 0.299& 0.900& 2.959\\
      \checkmark & \checkmark & $\bm{5.312}$& $\bm{0.296}$& $\bm{0.905}$& $\bm{3.718}$\\
     \midrule
     \multicolumn{2}{l}{\textbf{Tacotron 2}} & 5.770& 0.300 & 0.882& 3.409\\
    \bottomrule
  \end{tabular}
  \caption{Objective evaluation of score-based training, 100 inference steps, full validation (1800) set}
  \label{tab:loss}
\end{table}
The number of inference steps, $N$, for all EBMs was set to 100. Despite its simplicity, the delta loss trained EBM is competitive to the SSM trained EBM in all metrics other than UTMOSv2. The delta loss also appears to underperform compared to Tacotron 2 in the same metric while being superior based on all others. 
Combining SSM and delta losses does not appear to yield significant gains. 

Table\ref{tab:inference} shows how the number of inference steps affects objective evaluation metrics for SSM and delta loss trained EBMs. 
\begin{table}[htbp]
  \centering
  \begin{subtable}[h]{0.45\textwidth}
  \centering
  \begin{tabular}{c|cccc}
    \toprule
     \textbf{\scriptsize Step} & 
     \textbf{\scriptsize MCD$\downarrow$} &  
     \scriptsize $\log f_{\mathrm{o}} \downarrow$ &
     \textbf{\scriptsize S-BERT $\uparrow$} &
    \textbf{\scriptsize{UTMOSv2 $\uparrow$} } \\
    \midrule
     0 & 5.765& $\bm {0.297}$& 0.882& 3.412\\
     1 & $\bm{5.294}$& 0.299& $\bm {0.904}$& 3.717\\
     10 & 5.294& 0.299& 0.904& 3.708\\
     50 & 5.294& 0.300& 0.904& $\bm {3.719}$\\
     100 & 5.294& 0.300& 0.904& 3.704\\
    \bottomrule
  \end{tabular}
\subcaption{SSM}
\label{tab:inference ssm}
\end{subtable}
\hfill
\begin{subtable}[h]{0.45\textwidth}
  \centering
  \begin{tabular}{c|cccc}
    \toprule
     \textbf{\scriptsize Step} & 
     \textbf{\scriptsize MCD$\downarrow$} &  
     \scriptsize $\log f_{\mathrm{o}} \downarrow$ &
     \textbf{\scriptsize S-BERT $\uparrow$} &
    \textbf{\scriptsize{UTMOSv2 $\uparrow$} } \\
    \midrule
     0 & 5.765& 0.297& 0.882& 3.412\\
     1 & 5.292& 0.299& $\bm{0.904}$& $\bm{3.703}$\\
     10 & 5.277& 0.299& 0.904& 3.674\\
     50 & $\bm{5.273}$& $\bm{0.297}$& 0.903& 3.400\\
     100 & 5.389& 0.299& 0.900& 2.937\\
    \bottomrule
  \end{tabular}
    \caption{Delta}
  \label{tab:inference delta}
\end{subtable}
\caption{Objective evaluation of inference for score-based EBMs, partial validation (600) set}
\label{tab:inference}
\end{table}
These results suggest that for SSM loss the best results are obtained after just 1 step, which is in contrast to 100s of steps that may be required with NCE-trained EBMs and related diffusion models. Despite its simplicity, the delta loss appears to achieve similar results with also just 1 step. Although additional steps could improve some metrics they appear to lead to a consistent drop in UTMOSv2. 

Table~\ref{tab:inference with different loss} compares NCE~\cite{sun2024energy} and score-based trained EBMs. 
\begin{table}[htbp]
  \centering
  \begin{tabular}{cc|cccc}
    \toprule
     \textbf{\scriptsize Loss} & 
     \textbf{\scriptsize Step} &
     \textbf{\scriptsize MCD$\downarrow$} &  
     \scriptsize $\log f_{\mathrm{o}} \downarrow$ &
     \textbf{\scriptsize S-BERT $\uparrow$} &
    \textbf{\scriptsize{UTMOSv2 $\uparrow$} } \\
    \midrule
     $\text{NCE}^*$ & 100& 5.762& $\bm {0.295}$& 0.875& 2.572\\
     SSM & 1& 5.294& 0.299& 0.904& $\bm {3.717}$\\
     Delta & 1& $\bm {5.292}$& 0.299& 0.904& 3.703\\
    \bottomrule
  \end{tabular}
    \caption{Objective evaluation of NCE (full validation set) and score-based (partial validation set) trained EBMs}
  \label{tab:inference with different loss}
\end{table}
These results suggest that score-based training is a powerful alternative that yields significantly better MCD, S-BERT and UTMOSv2 values but may be insignificantly worse in $\log f_{\mathrm{o}}$.

\subsection{Subjective evaluation}
Finally, a range of listening tests was conducted to get an insight into score-based EBMs following the protocol outlined in Section~\ref{sec:evaluation}. Table~\ref{tab:subjective_metrics} summarises the conducted MOS study.
\begin{table}[htbp!]
\centering
\begin{tabular}{ ll|c }
\toprule
\# & \textbf{Method} & \textbf{MOS} \\
\midrule
1 & Ground Truth & 4.29$\pm$0.04 \\
2 & Ground Truth (log Mel + HiFi-GAN)& {\bf 4.15$\pm$0.06} \\
3 & Tacotron 2 (log Mel + HiFi-GAN)  & 3.74$\pm$0.08 \\
\midrule
4 & SSM (1 steps) (log Mel + HiFi-GAN) & 3.90$\pm$0.06\\
5 & Delta (1 steps) (log Mel + HiFi-GAN) & {\bf 3.95$\pm$0.05}\\
\midrule
6 & SSM (10 steps) (log Mel + HiFi-GAN) & 3.82$\pm$0.09\\
7 & Delta (10 steps) (log Mel + HiFi-GAN)& 3.76$\pm$0.07\\
\bottomrule
\end{tabular}
\caption{Subjective evaluation}
\label{tab:subjective_metrics}
\end{table}
The first block in Table~\ref{tab:subjective_metrics} shows MOS scores for original reference, ground truth, waveforms (line 1), ground truth waveforms resynthesised from log Mel spectrograms using HiFi-GAN vocoder (line 2) and the pretrained acoustic model, Tacotron 2, used in this paper (line 3). As expected, these results suggest that some level of degradation could be attributed to the imperfect vocoding process and that the pretrained acoustic model lags 0.41 in MOS behind that result. The second block confirms subjectively that the delta loss (line 5) proposed in this paper offers a computationally efficient powerful alternative to a more expensive and elaborate SSM loss (line 4). The delta loss trained EBM reduces the lag in MOS from resynthesised ground truth to 0.20. The final, third, block shows that although for SSM objective metrics appear to show little discrimination, the additional inference steps lead to a degradation in MOS. For delta loss, MOS and UTMOSv2 appear to be correlated. 

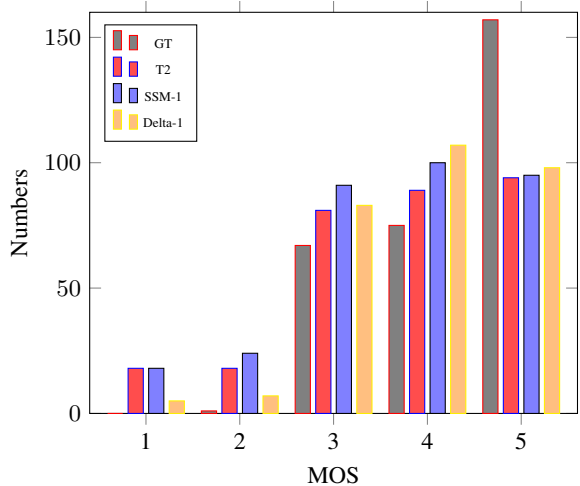
\begin{figure}[!htbp]
  \centering
  \begin{tikzpicture}
    \begin{axis}[
      ybar,
      bar width=0.2cm, 
      ymin=0,
      ymax=160, 
      symbolic x coords={1, 2, 3, 4, 5},
      xtick=data,
      xlabel={MOS},
      ylabel={Numbers},
      legend pos=north west,
      legend style={font=\tiny}, 
      enlarge x limits=0.15, 
      group style={group size=1 by 1, horizontal sep=1.2ex},
      cycle list name=color list,
      width=\columnwidth
    ]
      \addplot+[ybar, fill=black!50] coordinates {(1,0) (2,1) (3,67) (4,75) (5,157)};
      \addplot+[ybar, fill=red!70] coordinates { (1,18) (2,18) (3,81) (4,89) (5,94)};
      \addplot+[ybar, fill=blue!50] coordinates {(1,18) (2,24) (3,91) (4,100) (5,95)};
      \addplot+[ybar, fill=orange!50] coordinates {(1,5) (2,7) (3,83) (4,107) (5,98)};
      \legend{GT, T2, SSM-1, Delta-1}
    \end{axis}
  \end{tikzpicture}
  \caption{Detailed breakdown of MOS score counts}
  \label{tab:bar subjective_metrics}
\end{figure}

The breakdown of MOS score counts in Figure~\ref{tab:bar subjective_metrics} shows that although SSM and delta loss trained EBMs appear to be comparable in terms of the overall MOS scores, there are in fact important differences between these EBM systems. 
Figure~\ref{tab:bar subjective_metrics} suggests that the delta loss trained EBM has much fewer 1, 2 and 3 scores and more 4 and 5 scores than SSM trained EBM.

\section{Conclusion}
\label{sec:end}
This paper presented two new score-based approaches to training energy-based models (EBMs) for text-to-speech (TTS) as an alternative to more challenging Noise Contrastive Estimation (NCE). The first approach is sliced score matching (SSM) that has been previously used with EBMs in other areas but not in TTS. The second approach is a new objective function called {\em delta loss} which aims to promote log-likelihoods more suitable for first-order inference schemes used by EBMs and is simpler than NCE and SSM. This new loss was shown to be connected to losses used in currently popular flow matching.  
Experiments on the LJSpeech dataset show superior performance of score-based approaches compared to NCE. Despite its simplicity, the new delta loss function appears to show comparable performance in objective evaluations and better performance in subjective evaluations (less low scores and more high scores). 

\bibliographystyle{IEEEtran}
\bibliography{mybib}

\end{document}